\documentclass[twocolumn]{aastex631}

\usepackage{amsmath}
\usepackage{here}
\usepackage{xcolor}
\DeclareUnicodeCharacter{2212}{-}

\begin{document}

\shorttitle{
Strong Helium {\sc i} Emission Lines Found in High-$z$ Galaxies
}

\shortauthors{Yanagisawa et al.}

\title{
Strong He {\sc i} Emission Lines in High N/O Galaxies at $z\sim 6$ Identified in JWST Spectra:\\
High He/H Abundance Ratios or High Electron Densities?
}

\author[0009-0006-6763-4245]{Hiroto Yanagisawa}
\affiliation{Institute for Cosmic Ray Research, The University of Tokyo, 5-1-5 Kashiwanoha, Kashiwa, Chiba 277-8582, Japan}
\affiliation{Department of Physics, Graduate School of Science, The University of Tokyo, 7-3-1 Hongo, Bunkyo, Tokyo 113-0033, Japan}

\author[0000-0002-1049-6658]{Masami Ouchi}
\affiliation{National Astronomical Observatory of Japan, National Institutes of Natural Sciences, 2-21-1 Osawa, Mitaka, Tokyo 181-8588, Japan}
\affiliation{Institute for Cosmic Ray Research, The University of Tokyo, 5-1-5 Kashiwanoha, Kashiwa, Chiba 277-8582, Japan}
\affiliation{Department of Astronomical Science, SOKENDAI (The Graduate University for Advanced Studies), 2-21-1 Osawa, Mitaka, Tokyo, 181-8588, Japan}
\affiliation{Kavli Institute for the Physics and Mathematics of the Universe (WPI), University of Tokyo, Kashiwa, Chiba 277-8583, Japan}

\author[0000-0002-2740-3403]{Kuria Watanabe}
\affiliation{Department of Astronomical Science, SOKENDAI (The Graduate University for Advanced Studies), 2-21-1 Osawa, Mitaka, Tokyo, 181-8588, Japan}
\affiliation{National Astronomical Observatory of Japan, National Institutes of Natural Sciences, 2-21-1 Osawa, Mitaka, Tokyo 181-8588, Japan}

\author{Akinori Matsumoto}
\affiliation{Institute for Cosmic Ray Research, The University of Tokyo, 5-1-5 Kashiwanoha, Kashiwa, Chiba 277-8582, Japan}
\affiliation{Department of Physics, Graduate School of Science, The University of Tokyo, 7-3-1 Hongo, Bunkyo, Tokyo 113-0033, Japan}

\author[0000-0003-2965-5070]{Kimihiko Nakajima}
\affiliation{National Astronomical Observatory of Japan, National Institutes of Natural Sciences, 2-21-1 Osawa, Mitaka, Tokyo 181-8588, Japan}

\author[0000-0002-1319-3433]{Hidenobu Yajima}
\affiliation{Center for Computational Sciences, University of Tsukuba, Ten-nodai, 1-1-1 Tsukuba, Ibaraki 305-8577, Japan}

\author[0000-0001-7457-8487]{Kentaro Nagamine}
\affiliation{Theoretical Astrophysics, Department of Earth and Space Science, Graduate School of Science, Osaka University, Toyonaka, Osaka 560-0043, Japan}
\affiliation{Kavli Institute for the Physics and Mathematics of the Universe (WPI), University of Tokyo, Kashiwa, Chiba 277-8583, Japan}
\affiliation{Department of Physics \& Astronomy, University of Nevada, Las Vegas, 4505 S. Maryland Pkwy, Las Vegas, NV 89154-4002, USA}
\affiliation{Theoretical Joint Research, Forefront Research Center, Graduate School of Science, Osaka University, Toyonaka, Osaka, 560-0043, Japan}

\author[0000-0002-6705-6303]{Koh Takahashi}
\affiliation{National Astronomical Observatory of Japan, National Institutes of Natural Sciences, 2-21-1 Osawa, Mitaka, Tokyo 181-8588, Japan}

\author[0009-0000-1999-5472]{Minami Nakane}
\affiliation{Institute for Cosmic Ray Research, The University of Tokyo, 5-1-5 Kashiwanoha, Kashiwa, Chiba 277-8582, Japan}
\affiliation{Department of Physics, Graduate School of Science, The University of Tokyo, 7-3-1 Hongo, Bunkyo, Tokyo 113-0033, Japan}

\author[0000-0001-8537-3153]{Nozomu Tominaga}
\affiliation{National Astronomical Observatory of Japan, National Institutes of Natural Sciences, 2-21-1 Osawa, Mitaka, Tokyo 181-8588, Japan}
\affiliation{Department of Physics, Faculty of Science and Engineering, Konan University, 8-9-1 Okamoto, Kobe, Hyogo 658-8501, Japan}
\affiliation{Department of Astronomical Science, SOKENDAI (The Graduate University for Advanced Studies), 2-21-1 Osawa, Mitaka, Tokyo, 181-8588, Japan}

\author[0009-0008-0167-5129]{Hiroya Umeda}
\affiliation{Institute for Cosmic Ray Research, The University of Tokyo, 5-1-5 Kashiwanoha, Kashiwa, Chiba 277-8582, Japan}
\affiliation{Department of Physics, Graduate School of Science, The University of Tokyo, 7-3-1 Hongo, Bunkyo, Tokyo 113-0033, Japan}

\author[0000-0002-0547-3208]{Hajime Fukushima}
\affiliation{Center for Computational Sciences, University of Tsukuba, Ten-nodai, 1-1-1 Tsukuba, Ibaraki 305-8577, Japan}

\author[0000-0002-6047-430X]{Yuichi Harikane}
\affiliation{Institute for Cosmic Ray Research, The University of Tokyo, 5-1-5 Kashiwanoha, Kashiwa, Chiba 277-8582, Japan}

\author[0000-0001-7730-8634]{Yuki Isobe}
\affiliation{Waseda Research Institute for Science and Engineering, Faculty of Science and Engineering, Waseda University, 3-4-1, Okubo, Shinjuku, Tokyo 169-8555, Japan
}

\author[0000-0001-9011-7605]{Yoshiaki Ono}
\affiliation{Institute for Cosmic Ray Research, The University of Tokyo, 5-1-5 Kashiwanoha, Kashiwa, Chiba 277-8582, Japan}

\author[0000-0002-5768-8235]{Yi Xu}
\affiliation{Institute for Cosmic Ray Research, The University of Tokyo, 5-1-5 Kashiwanoha, Kashiwa, Chiba 277-8582, Japan}
\affiliation{Department of Astronomy, Graduate School of Science, The University of Tokyo, 7-3-1 Hongo, Bunkyo, Tokyo 113-0033, Japan}

\author[0000-0003-3817-8739]{Yechi Zhang}
\affiliation{National Astronomical Observatory of Japan, National Institutes of Natural Sciences, 2-21-1 Osawa, Mitaka, Tokyo 181-8588, Japan}

\begin{abstract}
We present He\textsc{i}/H$\beta$-flux and He/H-abundance ratios in three JWST galaxies with significant constraints on N/O-abundance ratios, GS-NDG-9422, RXCJ2248-ID, and GLASS150008 at $z\sim 6$ mostly with the spectroscopic coverage from He\textsc{i}$\lambda$4471 and He\textsc{ii}$\lambda$4686 to He\textsc{i}$\lambda$7065, comparing with 68 local-dwarf galaxies. We find that these high-$z$ galaxies present strong He\textsc{i} emission with He\textsc{i}/H$\beta$ flux ratios generally larger than those of local-dwarf galaxies. 
We derive He/H with all of the detected He\textsc{i}, He\textsc{ii}, and $2-3$ hydrogen Balmer lines in the same manner as the local He/H determination conducted for cosmology studies. These high-$z$ galaxies show He overabundance He/H$\gtrsim 0.10$ or high electron density $n_{\rm e}\sim 10^{3-4}$ cm$^{-3}$ much larger than local values at low O/H, $12+\log{\rm (O/H)}=7-8$.
In contrast, we obtain low He/H and $n_{\rm e}$ values for our local-dwarf galaxies by the same technique with the same helium and hydrogen lines, and confirm that the difference between the high-$z$ and local-dwarf galaxies are not mimicked by systematics.
While two scenarios of 1) He overabundance and 2) high electron density are not clearly concluded, we find that there is a positive correlation on the He/H-N/O or $n_{\rm e}$-N/O plane by the comparison of the high-$z$ and local-dwarf galaxies. The scenario 1) suggests that the overabundant helium and nitrogen are not explained by the standard chemical enrichment of core-collapse supernovae, but the CNO-cycle products and equilibrium ratios, respectively. The scenario 2) indicates that the strong helium lines are originated from the central dense clouds of the high-$z$ galaxies by excessive collisional excitation.
\end{abstract}


\section{Introduction} \label{sec:intro}
The James Webb Space Telescope (JWST) provides high-sensitivity near-infrared (NIR) spectroscopy and allows us to detect various elemental emission lines of high-$z$ galaxies.
The high N/O values in high-$z$ galaxies such as GN-z11 are reported in recent studies \citep{Cameron+2023a, Senchyna+2023, Isobe+2023c}.
\cite{Isobe+2023c} investigate abundances of C, N, and O in high-$z$ galaxies and suggest that the low C/N and high N/O ratios are explained by the chemical composition of CNO-cycle equilibrium. \cite{Isobe+2023c} also find that the low C/N and high N/O ratios found in the high-$z$ galaxies coincide with those of globular cluster stars, suggesting that these galaxies may be forming globular clusters.
\cite{Topping+2024} present RXCJ2248-ID at $z\sim6$ showing not only high N/O but also high electron densities, and suggest that CNO-cycled gas coincide with the formation of globular clusters under the very dense conditions. This may be connected with a trend of high electron densities in high-$z$ galaxies, as suggested by \cite{Isobe+2023_ne}.

Because the CNO-cycle produces not only a large amount of nitrogen but also helium, the high-$z$ galaxies that have the high N/O abundances may show high He/H abundances. Strong He \textsc{i} emission lines are detected in some of the high-$z$ galaxies \citep{Cameron+2023, Topping+2024}. 
In this work, we clarify whether there is a difference in helium abundance between local and high-$z$ galaxies by comparing intensities of the He \textsc{i} emission lines. We also investigate the connection between He/H, N/O and the electron density.
We derive helium abundances with high accuracy using a method developed for Big-Bang nucleosynthesis studies \citep{Izotov+2014, Aver+2015, Hsyu+2020, Matsumoto+2022}. This work aims to understand the process of chemical enrichment in the high-$z$ universe by comparing the observed helium abundances with local galaxies and chemical evolution models. 

This paper is organized as follows. In Section \ref{sec:data}, we present our sample of galaxies. In Section \ref{sec:chemical_abundance}, we derive the helium abundances of the galaxies in our sample. Section \ref{sec:result} shows our results of He/H measurements and the correlation between the He/H and N/O ratios. We also present the correlation between the electron densities and N/O in Section \ref{sec:result}. We compare our observational results with several chemical evolution models in Section \ref{sec:discussion}. We summarize our results in Section \ref{sec:summary}.

\section{Data and Sample} \label{sec:data}
At present there are 5 galaxies at $z \gtrsim 6$ whose N/O values are significantly constrained (GN-z11; \citealt{Cameron+2023a, Charbonnel+2023}; CEERS01019, GLASS150008; \citealt{Isobe+2023c}, RXCJ2248-ID; \citealt{Topping+2024}, GS-NDG-9422; \citealt{Cameron+2023}). To derive He/H, hydrogen and helium emission lines within the rest-frame wavelength range approximately from $4000 \, \mathrm{\AA}$ to $7000 \, \mathrm{\AA}$ are needed, which requires galaxies to be in the redshift range of $0.5 \lesssim z \lesssim 6.5$. We select 3 galaxies (GS-NDG-9422, RXCJ2248-ID, and GLASS150008) that satisfy these criteria, which are presented in Sections \ref{sec:gs}, \ref{sec:rxc}, and \ref{sec:glass}. Note that we do not require large N/O values in order to obtain the unbiased sample.
Local dwarf galaxies are also presented as a control sample in Section \ref{sec:local}.

\begin{deluxetable*}{ccccc}
    \centering
    \tablecaption{Observed emission line flux ratios}
    \tablehead{
    \colhead{Emission line} & \colhead{GS-NDG-9422} & \colhead{RXCJ2248-ID} & \colhead{GLASS150008} & \colhead{J2115-1734$^\dagger$}
    }
    \startdata
    H$\delta$                    & $25.0 \pm 2.9$ & $28.3 \pm 1.2$ & $25.4 \pm 2.5$ & $27.0\pm0.2$ \\
    H$\gamma$                    & $48.3 \pm 2.3$ & $51.2 \pm 0.8$ & $\dots$ & $47.1\pm0.3$ \\
    He \textsc{i} $\lambda$4471  & $4.7 \pm 0.6$ & $9.4 \pm 0.8$ & $\dots$ & $3.8\pm0.1$ \\
    He \textsc{ii} $\lambda$4686 & $6.4\pm1.2$ & $\dots$ & $\dots$ & $2.6\pm0.2$ \\
    H$\beta$                     & $100.0 \pm 1.7$ & $100.0 \pm 0.8$ & $100.0 \pm 3.7$ & $100.0\pm0.4$ \\
    He \textsc{i} $\lambda$5876  & $12.8 \pm 0.58$ & $20.5 \pm 0.8$ & $18.7 \pm 3.2$ & $10.9\pm0.1$ \\
    H$\alpha$                    & $264.5 \pm 3.5$ & $254.7 \pm 0.8$ & $\dots$ & $288.9\pm0.1$ \\
    He \textsc{i} $\lambda$7065  & $7.0\pm0.6$ & $12.2 \pm 2.0$ & $\dots$ & $3.7\pm0.2$ \\
    \enddata
    \tablecomments{Observed flux ratios relative to H$\beta$ ($\times 100$). 
    For GS-NDG-9422 and RXCJ2248-ID, the flux values are taken from \cite{Topping+2024} and \cite{Cameron+2023}, respectively. ($\dagger$) One out of the 68 local galaxies is presented as an example.
    }
    \label{tab:flux}
\end{deluxetable*}

\begin{figure*}
    \centering
    \includegraphics[width=1\linewidth]{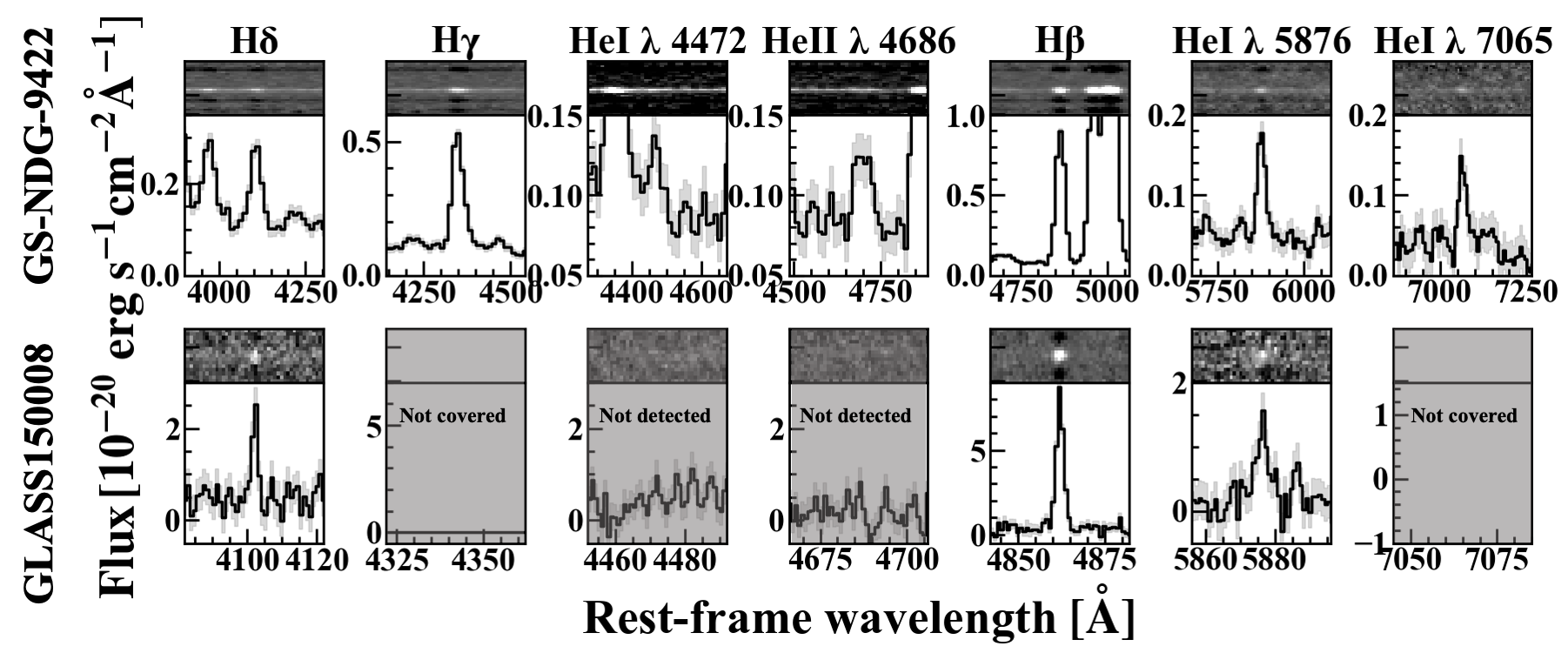}
    \caption{2D and 1D spectra of  GS-NDG-9422 (top) and GLASS150008 (bottom). The black solid lines and gray shaded regions represent the spectra and 1$\sigma$ errors, respectively. From left to right, H$\delta$, H$\gamma$, He \textsc{i} 4472, He \textsc{ii} $\lambda$4686, H$\beta$, He \textsc{i} $\lambda$5876, and He \textsc{i} $\lambda$7065 are shown. The undetected emission lines are flagged as `Not detected'. The emission lines that are out of the wavelength range of the detector are flagged as `Not covered'.}
    \label{fig:spectra}
\end{figure*}

\subsection{GS-NDG-9422} \label{sec:gs}
Spectroscopic data of GS-NDG-9422 at $z=5.94$ was obtained with JWST NIRSpec via the JWST Advanced Deep Extragalactic Survey (JADES; PID: 1210, PI: Luetzgendorf; \citealt{Bunker+2023, Eisenstein+2023}). \cite{Cameron+2023} have investigated the spectra and measured the rest-UV and optical emission line fluxes. We calculate the flux ratios relative to H$\beta$ using the flux values of prism spectrum reported in \cite{Cameron+2023}. The flux ratios are shown in Table \ref{tab:flux}. Since the prism spectrum covers all of the emission lines used in this study, systematic uncertainty caused by slit-loss is negligibly small. The prism spectrum of GS-NDG-9422 is shown in Figure \ref{fig:spectra}. 

Since no nitrogen line is detected in GS-NDG-9422, \cite{Cameron+2023} derived 3$\sigma$ upper limits of $\log (\mathrm{N/O}) < -0.85 $ and $-1.01$ using N \textsc{iii}]$\lambda$1750 / [O \textsc{iii}]$\lambda$5007 and N \textsc{iii}]$\lambda$1750 / [O \textsc{iii}]$\lambda$1666, respectively. These N/O ratios are consistent with those of local galaxies. In this study, we adopt the more conservative upper limit of $\log (\mathrm{N/O}) < -0.85$.

\subsection{RXCJ2248-ID} \label{sec:rxc}
JWST NIRSpec observation of RXCJ2248-ID at $z=6.11$ was conducted in General Observers (GO) program (PID: 2478, PI: Stark). \cite{Topping+2024} have analysed the spectra and measured the rest-UV and optical emission line fluxes. The flux ratios relative to H$\beta$ are calculated with the flux values reported in \cite{Topping+2024}. These flux ratios are shown in Table \ref{tab:flux}.

\subsection{GLASS150008} \label{sec:glass}
GLASS150008 at $z=6.23$ was observed with the JWST NIRSpec in the GLASS JWST Early Release Science (GLASS; Proposal ID: 1324, PI: T. Treu; \citealt{Treu+2022}). The GLASS 
data were acquired by using high resolution ($R \sim 2700$) filter-grating pairs of F100LP-G140H, F170LP-G235H, and F290LP-G395H, which cover the wavelength ranges of $1.0−1.6$, $1.7−3.1$ and $2.9−5.1$ $\mu$m, respectively. The total exposure time of the GLASS data is 4.9 hours for each filter-grating pair. We use the spectroscopic data that were reduced by \cite{Nakajima+2023}. \cite{Nakajima+2023} conducted the level-2 and -3 calibrations on the raw data using the JWST Science Calibration Pipeline version 1.8.5 \citep{jwst_pipeline_1_8_5} with the Calibration Reference Data System (CRDS) context file of $\tt\string jwst\_1028.pmap$ or $\tt\string jwst\_1027.pmap$. \cite{Nakajima+2023} also performed additional processes that improve the flux calibration, noise estimate, and composition. See \cite{Nakajima+2023} for the details of the data reduction. The reduced spectrum of GLASS150008 is shown in Figure \ref{fig:spectra}.

\cite{Isobe+2023c} have derived the N/O ratio using the N \textsc{iii}]$\lambda$1750 / O \textsc{iii}]$\lambda\lambda$1661,1666 and obtained $\log (\mathrm{N/O}) = -0.40^{+0.05}_{-0.07}$, which is significantly higher than those of typical local galaxies.

We measure flux values of the hydrogen and helium emission lines that are used for helium abundance estimation by fitting a Gaussian function. We detect H$\delta$, H$\beta$, and He \textsc{i} $\lambda$ 5876 at more than the 6$\sigma$ levels. We do not use He \textsc{i} $\lambda$3889 and 5015 because these lines are blended with H8 and [O {\sc iii}] $\lambda$5007, respectively. The errors of fluxes are obtained from uncertainties of the Gaussian fit. The measured emission line fluxes are listed in Table \ref{tab:flux}.

\begin{deluxetable*}{ccccc}
    \centering
    \tablecaption{Summary of our sample}
    \tablehead{ 
    \colhead{Name} & \colhead{$z$} & \colhead{$\log M_*/M_\odot$} & \colhead{$\log \mathrm{SFR}/M_\odot \, \mathrm{yr^{-1}}$} & Ref. \\
    (1) & (2) & (3) & (4) & (5)
    }
    \startdata
    GS-NDG-9422 & 5.94 & 
    $\mathbf{7.80^{{+0.01}^\dagger}_{-0.01}}$ & $0.91^{+0.13}_{-0.18}$ & 
    \cite{Terp+2024} \\
    RXCJ2248-ID & 6.11 & $8.05^{+0.17}_{-0.15}$ & $1.8^{+0.2}_{-0.2}$ & \cite{Topping+2024} \\
    GLASS150008 & 6.23 & $8.39^{+0.35}_{-0.19}$ & $1.0^{+0.9}_{-0.7}$ & \cite{Jones+2023}\\
    \tableline 
    J2115-1734$^\ddagger$ & 0.02 & $6.56^{+0.02}_{-0.02}$ & $0.27^{+0.01}_{-0.01}$  &  \cite{Kojima+2020} \\
    \enddata
    \tablecomments{(1) Name. (2) Redshift. (3) Stellar mass. (4) Star-formation rate. (5) Reference for the stellar mass and star-formation rate. 
    ($\dagger$) While \cite{Terp+2024} assume stellar continuum for GS-NDG-9422 and present $\log M_*/\mathrm{M_\odot} = 7.80^{+0.01}_{-0.01}$, \cite{Cameron+2023} assume that the nebular continuum is dominant and does not present the stellar mass.
    ($\ddagger$) One out of the 68 local galaxies is presented as an example.}
    \label{tab:sample}
\end{deluxetable*}

\subsection{Local Dwarf Galaxies} \label{sec:local}
As a control sample, we use a total of 68 local dwarf galaxies reported by \cite{Hsyu+2020}, \cite{Matsumoto+2022} and Yanagisawa et al. (in prep.), which have He/H and O/H measurements and O/H lower than 0.4 $Z_\odot$. These galaxies have stellar masses of $\log M_*/\mathrm{M_\odot} \sim 6.5-10$ and star-formation rates (SFRs) of $\log \mathrm{SFR/M_\odot yr^{-1}} \sim 0-1$.
For the galaxies reported by \cite{Hsyu+2020} and \cite{Matsumoto+2022}, we use the flux values that are presented by \cite{Hsyu+2020} and \cite{Matsumoto+2022}, respectively. For the other galaxies that will be reported by Yanagisawa et al. (in prep.), optical spectra are obtained from SDSS and by Magellan/MagE observation (PI: M. Rauch). For detail of the MagE observation, see \cite{Xu+2022}. Emission line fluxes and errors are measured in the same manner as explained in Section \ref{sec:glass}. The calculated flux ratios for one out of the 68 local galaxies, J2115-1734, are presented in Table \ref{tab:flux}.

\subsection{Our Sample}
\begin{figure}
    \centering
    \includegraphics[width=1\linewidth]{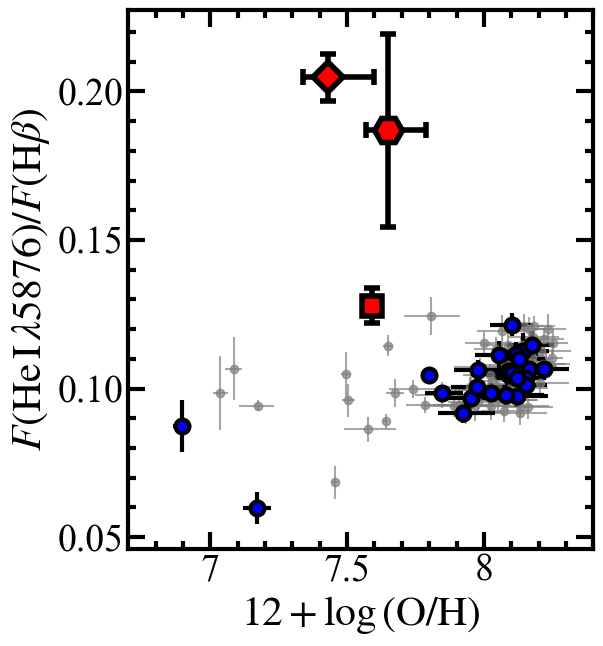}
    \caption{Flux ratios of He \textsc{i} $\lambda$5876/H$\beta$ as a function of O/H. The red square, diamond, and hexagon represent GS-NDG-9422, RXCJ2248-ID, and GLASS150008, respectively. The blue and gray dots denote the local galaxies, where the blue dots are the galaxies that have N/O constraints.}
    \label{fig:fluxratio}
\end{figure}

Our sample comprises of a total of 3 high-$z$ galaxies and 68 local galaxies. Hereafter these 71 ($=3+68$) galaxies are referred to as our sample. We also summarize the 3 high-$z$ galaxies and 1 representative local galaxy in Table \ref{tab:sample}. Figure \ref{fig:fluxratio} presents the He \textsc{i} $\lambda$5876/H$\beta$ ratios of these galaxies. RXCJ2248-ID and GLASS150008 show significantly intense He \textsc{i} emission compared to those of the local galaxies, while GS-NDG-9422 shows the He \textsc{i} emission comparable to (but moderately stronger than) those of the local galaxies. 

\section{Chemical Abundances} \label{sec:chemical_abundance}

\subsection{Helium Abundance} \label{sec:he_abundance}
The number abundance ratio of helium to hydrogen $y \equiv {\rm He/H}$ is given by

\begin{equation}
    y = \frac{\rm He^{+}}{\rm H^{+}} + \frac{\rm He^{++}}{\rm H^{+}} = y^{+} + y^{++},
\end{equation}
where $y^{+}$ and $y^{++}$ are the number abundance ratios of singly and doubly ionized helium to ionized hydrogen, respectively. We ignore the neutral hydrogen and helium because the observed emission lines come from H {\sc ii} regions, where most of the hydrogen and helium atoms are ionized.

We present the methods to obtain $y^{+}$ and $y^{++}$ in sections \ref{sec:y+} and \ref{sec:y++}, respectively.

\begin{figure*}
    \centering
    \includegraphics[width=1\linewidth]{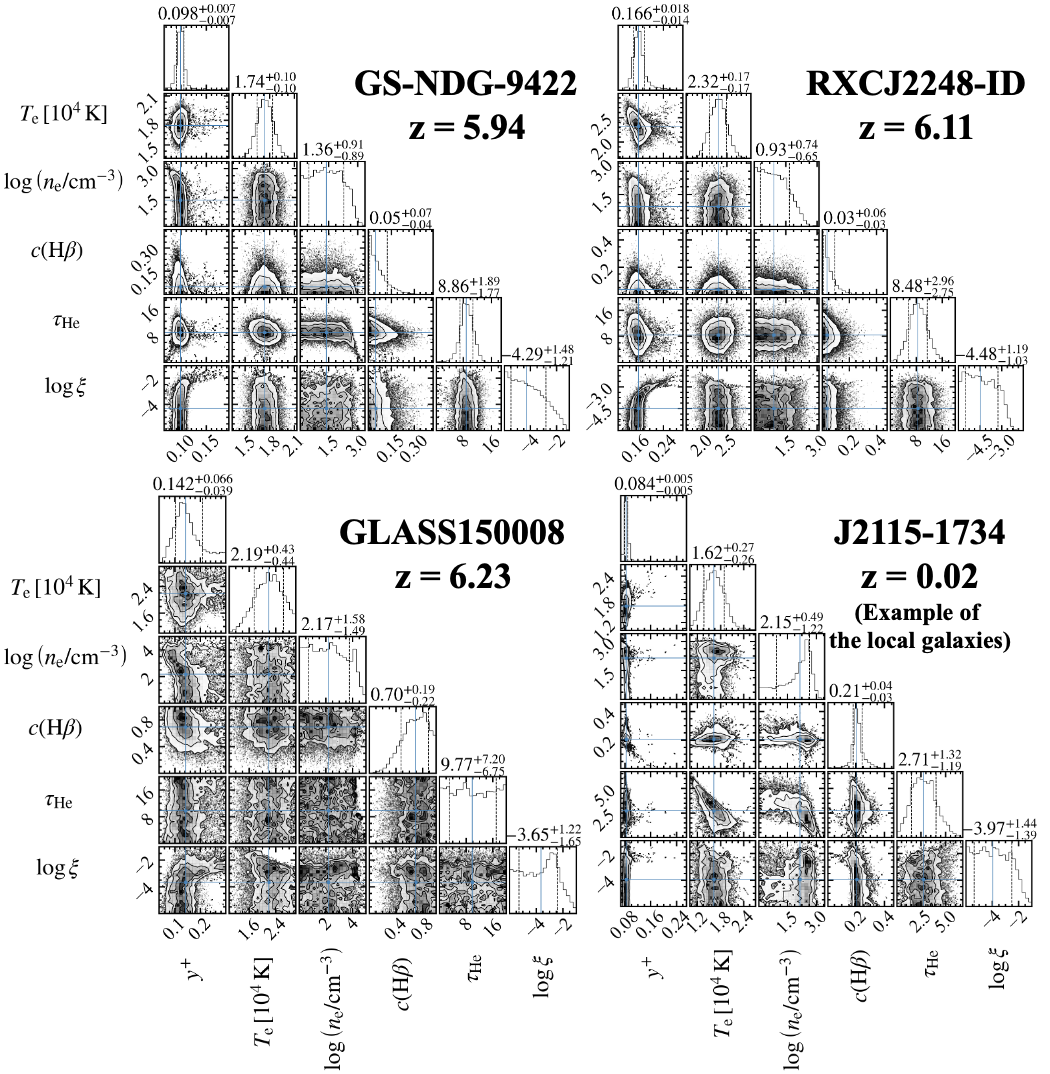}
    \caption{Probability distribution functions (PDFs) obtained with modified YMCMC. The diagonal and off-diagonal panels show 1D and 2D PDFs, respectively. The blue solid lines and black dashed lines represent the median and 68\% confidence levels, respectively. On top of each column, the median and 68\% confidence level values of the PDF are presented.}
    \label{fig:mcmc}
\end{figure*}

\begin{deluxetable*}{cccccccc}
    \tablewidth{12pt}
    \tablecaption{Best recovered parameters from MCMC analysis} 
    \tablehead{ 
    \colhead{Name} & \colhead{$y^{+}$} & \colhead{$T_\mathrm{e}$} & \colhead{$\log (n_\mathrm{e}/\mathrm{cm^{-3}})$} & \colhead{$c$(H$\beta$)} & \colhead{$\tau_{\rm He}$} & \colhead{$\log\xi$} & \colhead{$\log N_2$} \\
    &&[K]&&&&& \\
    (1)&(2)&(3)&(4)&(5)&(6)&(7)&(8)
    }
    \startdata
    GS-NDG-9422 & $0.098^{+0.007}_{-0.007}$ & $17379^{+1006}_{-955}$ & $1.36^{+0.91}_{-0.89}$ & $0.05^{+0.07}_{-0.04}$ & $8.86^{+1.89}_{-1.77}$ & $-4.3^{+1.5}_{-1.2}$ & $12.5^{+0.3}_{-0.2}$ \\
    RXCJ2248-ID & $0.166^{+0.018}_{-0.014}$ & $23241^{+1711}_{-1686}$ & $0.93^{+0.74}_{-0.65}$ & $0.03^{+0.06}_{-0.03}$ & $8.48^{+2.96}_{-2.75}$ & $-4.5^{+1.2}_{-1.0}$ & $13.4^{+0.1}_{-0.1}$ \\
    GLASS150008 & $0.142^{+0.066}_{-0.039}$ & $21889^{+4320}_{-400}$ & $2.17^{+1.58}_{-1.49}$ & $0.70^{+0.19}_{-0.22}$ & $\dots^\dagger$ & $-3.7^{+1.2}_{-1.7}$ & $11.2^{+1.5}_{-1.5}$ \\
    \tableline
    J2115-1734$^\ddagger$ & $0.084^{+0.005}_{-0.005}$ & $16238^{+2703}_{-2553}$ & $2.15^{+0.49}_{-1.22}$ & $0.21^{+0.04}_{-0.03}$ & $2.71^{+1.32}_{-1.19}$ & $-4.0^{+1.4}_{-1.5}$ & $11.4^{+1.4}_{-1.5}$ \\
    \enddata
    \tablecomments{(1) Name. (2) Number abundance ratio of singly ionized helium to hydrogen. (3) Electron temperature. (4) Electron density. (5) Reddening correction coefficient. (6) Helium optical depth parameter. ($\dagger$) For GLASS150008, best-fit $\tau_\mathrm{He}$ is not shown because the obtained probability distribution is almost constant. (7) Number abundance ratio of neutral to ionized hydrogen. (8) Column density of the hydrogen atom at the state with the principal quantum number 2. ($\ddagger$) One out of the 68 local galaxies is presented as an example.} 
    \label{table:mcmc_recovered_params}
\end{deluxetable*}

\begin{deluxetable*}{ccccccc}
    \tablewidth{12pt}
    \tablecaption{Chemical abundances} 
    \tablehead{ 
    \colhead{Name} & \colhead{$12+\log(\rm O/\rm H)$} & \colhead{$\log (\mathrm{N/O})$} & \colhead{$y^{+}$} & \colhead{$y^{++}$} & \colhead{$y$} & Ref.\\
    (1)&(2)&(3)&(4)&(5)&(6)&(7)
    }
        \startdata
        GS-NDG-9422 & $7.59^{+0.01}_{-0.01}$ & $<-0.85$ & $0.098^{+0.007}_{-0.007}$ & $0.006^{+0.001}_{-0.001}$ & $0.104^{+0.007}_{-0.007}$ & \cite{Cameron+2023}\\
        RXCJ2248-ID & $7.43^{+0.17}_{-0.09}$ & $-0.39^{+0.10}_{-0.08}$ & $0.166^{+0.018}_{-0.014}$ & $<0.005$ & $0.166^{+0.018}_{-0.014}$ & \cite{Topping+2024}\\
        GLASS150008 & $7.65^{+0.14}_{-0.08}$& $-0.40^{+0.05}_{-0.07}$ & $0.142^{+0.066}_{-0.039}$ & $<0.002$ & $0.142^{+0.066}_{-0.039}$ & \cite{Isobe+2023c}\\
        \tableline
        J2115-1734$^\dagger$ & $4.27^{+0.10}_{-0.10}$ & $-1.56^{+0.01}_{-0.01}$ & $0.084^{+0.005}_{-0.005}$ & $0.0023^{+0.0002}_{-0.0002}$ & $0.086^{+0.005}_{-0.005}$ & \cite{Kojima+2021}\\
        \enddata
    \tablecomments{(1) Name. (2) Oxygen abundance. (3) N/O abundance ratio. (4) Singly ionized helium abundance. (5) Doubly ionized helium abundance. (6) Total helium abundance. (7) Reference for the O/H and N/O ratios. ($\dagger$) One out of the 68 local galaxies is presented as an example.} 
    \label{table:abundance}
\end{deluxetable*}

\subsubsection{Singly Ionized Helium} \label{sec:y+}
The $y^{+}$ value is estimated with modified YMCMC, which is based on YMCMC \citep{Hsyu+2020} and modified to model the optical depth for the Balmer lines by \cite{Yanagisawa+2024}. Modified YMCMC uses the Markov Chain Monte Carlo (MCMC) algorithm and searches for the best-fit parameters that reproduce the observed hydrogen and helium emission lines (for detail, see \citealt{Hsyu+2020} and \citealt{Yanagisawa+2024}). We constrain 7 free parameters of $y^{+}$, the electron temperature $T_\mathrm{e}$, electron density $n_{\rm e}$, reddening correction parameter $c({\rm H\beta})$, helium optical depth $\tau_{\rm He}$, neutral to ionized hydrogen fraction $\xi$, and column density of the hydrogen atom with a principal quantum number 2 $N_2$.
We use the flat priors of

\begin{equation}\label{eq:prior}
\begin{split}
    &0.01 \leq y^+ \leq 0.30, \\
    &1 \leq \log_{10}(n_\mathrm{e}/\mathrm{cm^{-3}}) \leq 5, \\
    &0 \leq c({\rm H\beta}) \leq 1, \\
    &0 \leq \tau_\mathrm{He} \leq 20, \\
    &-6 \leq \log_{10}(\xi) \leq -1, \\
    &9 \leq \log_{10}(N_2/\mathrm{cm^{-2}}) \leq 15.
\end{split}
\end{equation}
We also use a Gaussian prior for $T_\mathrm{e}$ to prevent the walkers from falling in an unphysical solution, following the method presented by \citep{Aver+2010}. We use $T_{\rm e}(\textrm{He} \, \textsc{ii}) = 0.95 \, T_{\rm e}(\textrm{O} \, \textsc{iii})$ as a central value of a Gaussian prior, where $T_{\rm e}(\textrm{He} \, \textsc{ii})$ and $T_{\rm e}(\textrm{O} \, \textsc{iii})$ is the electron temperature of a He {\sc ii} and an [O {\sc iii}] region, respectively. This assumption is based on \cite{Peimbert+2002}, who claim that $T_\mathrm{e}(\mathrm{He} \, \textsc{ii})$ is 0.9-1.0 times lower than $T_{\rm e}(\textrm{O} \, \textsc{iii})$ in typical H \textsc{ii} regions. The obtained probability distributions for GS-NDG-9422, RXCJ2248-ID, and GLASS150008 are presented in Figure \ref{fig:mcmc}. These high-$z$ galaxies show $y^{+} \gtrsim 0.10$, which is remarkably larger than those of the local galaxies ($y^{+} \sim 0.08$; \citealt{Hsyu+2020, Matsumoto+2022}). The probability distributions for GS-NDG-9422 and RXCJ2248-ID present $\tau_\mathrm{He} \sim 8-9$, which are significantly larger than those of typical local galaxies ($\tau_\mathrm{He} \lesssim 3$; \citealt{Hsyu+2020, Matsumoto+2022}). This is likely to coincide with the optically-thick hydrogen gas suggested by \cite{Yanagisawa+2024}. We also derive He/H for the control sample of the local galaxies in the same manner, using the same helium and hydrogen lines as the input. The probability distribution of one of the local galaxies, J2115-1734, is also shown in Figure \ref{fig:mcmc}. The $y^{+}$ value for J2115-1734 is comparable to the typical values of 0.08, indicating that the limited number of the input emission lines is not the cause of the large $y^{+}$ in the high-$z$ galaxies. We also run the YMCMC code that is developed by \cite{Hsyu+2020} in the same manner, and confirm that the results do not change.

\subsubsection{Doubly Ionized Helium} \label{sec:y++}
We derive the $y^{++}$ values from Equation (17) of \cite{Pagel+1992}:

\begin{equation}
    y^{++} = 0.084 \left\lbrack\frac{T_{\textrm{e}}(\textrm{O\,\textsc{iii}})}{10^4}\right\rbrack^{0.14}\frac{F\left(\textrm{He\,\textsc{ii}}\,\lambda4686\right)}{F\left(\textrm{H}\beta\right)} ,
\end{equation}
where $F(\lambda)$ represents the flux value of the emission line $\lambda$. 
Measurements and errors of $T_{\rm e}(\textrm{O} \, \textsc{iii})$ are taken from \citet{Cameron+2023}, \cite{Topping+2024}, and 
\citet{Jones+2023} for GS-NDG9422, RXCJ2248-ID, and GLASS150008, respectively. For the local galaxies presented by \cite{Hsyu+2020} and \cite{Matsumoto+2022}, $T_\mathrm{e}(\mathrm{O \, \textsc{iii}})$ values are taken from \cite{Hsyu+2020}. The $T_\mathrm{e}(\mathrm{O \, \textsc{iii}})$ values for the galaxies that will be presented by Yanagisawa et al. (in prep.) are taken from \cite{Kojima+2020}, \cite{Izotov+2020, Izotov+2021}, and \cite{SanchezAlmeida+2015}.
If the He {\sc ii} $\lambda$4686 line is not detected, we put the 2$\sigma$ upper limit on the $y^{++}$ value. We present the $y^{++}$ values of our sample galaxies in Table \ref{table:abundance}.

\subsection{Oxygen and Nitrogen Abundances} \label{sec:o_abundance}
The O/H and N/O abundance ratios of GS-NDG-9422, RXCJ2248-ID, and GLASS150008 are taken from \cite{Cameron+2023}, \cite{Topping+2024}, and \cite{Isobe+2023c}, respectively. The O/H abundance ratios of the local galaxies that are presented by \cite{Hsyu+2020} and \cite{Matsumoto+2022} are taken from \cite{Hsyu+2020} and \cite{Matsumoto+2022}, respectively. The O/H abundance ratios for the galaxies that will be presented by Yanagisawa et al. (in prep.) are taken from \cite{Kojima+2020}, \cite{Izotov+2020, Izotov+2021}, and \cite{SanchezAlmeida+2015}. The N/O abundance ratios of the local dwarf galaxies are taken from \cite{Hagele+2011}, \cite{Puertas+2022}, and \cite{Kojima+2021}. The O/H and N/O abundance ratios of the 3 high-$z$ galaxies and 1 representative local galaxy in our sample are presented in Table \ref{table:abundance}. 

\begin{figure}
    \centering
    \includegraphics[width=1\linewidth]{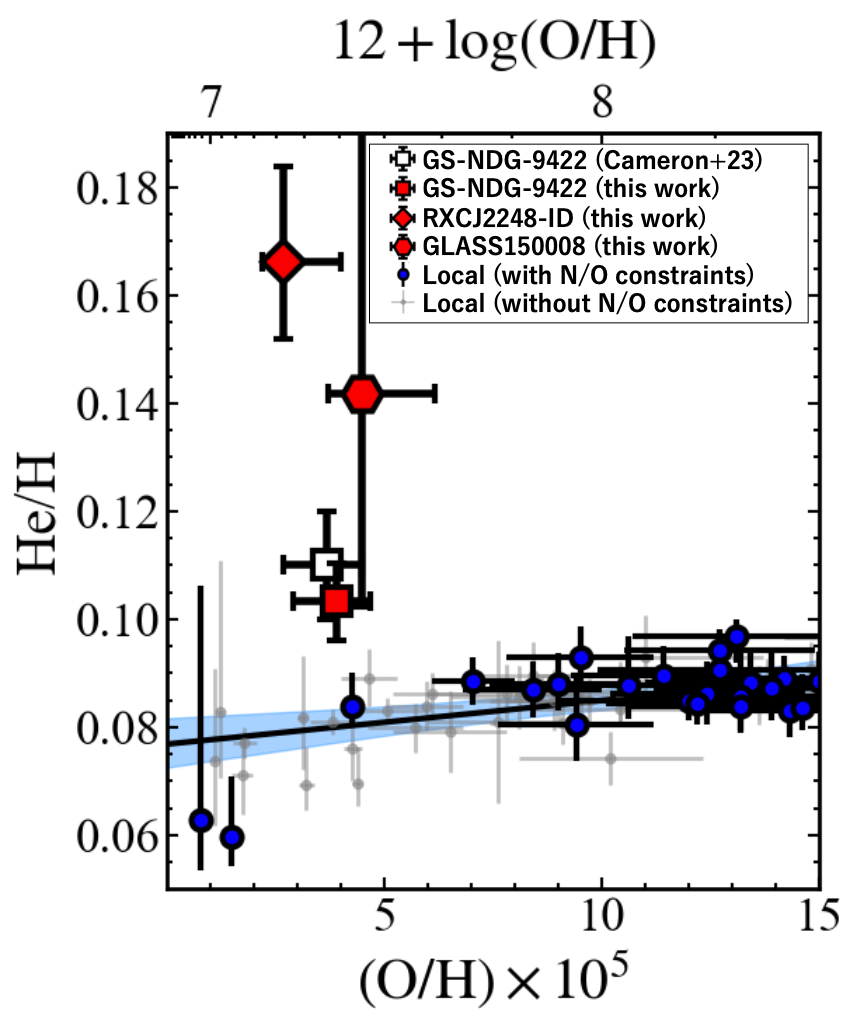}
    \caption{Helium abundance as a function of oxygen abundance. The squares, diamonds, and hexagons represent GS-NDG-9422, RXCJ2248-ID, and GLASS150008, respectively. The open square denotes the He/H derived by \cite{Cameron+2023}. To avoid overlapping the data points, the open square is shifted to the left by 0.2 dex. The blue and gray points denote the local dwarf galaxies (\citealt{Hsyu+2020}, \citealt{Matsumoto+2022}, and Yanagisawa et al. in prep.) with and without N/O measurement, respectively. The black line and blue shaded region are the linear relation of local galaxies and its 3$\sigma$ error, respectively \citep{Matsumoto+2022}.}
    \label{fig:HeH-OH}
\end{figure}

\begin{figure}
    \centering
    \includegraphics[width=1\linewidth]{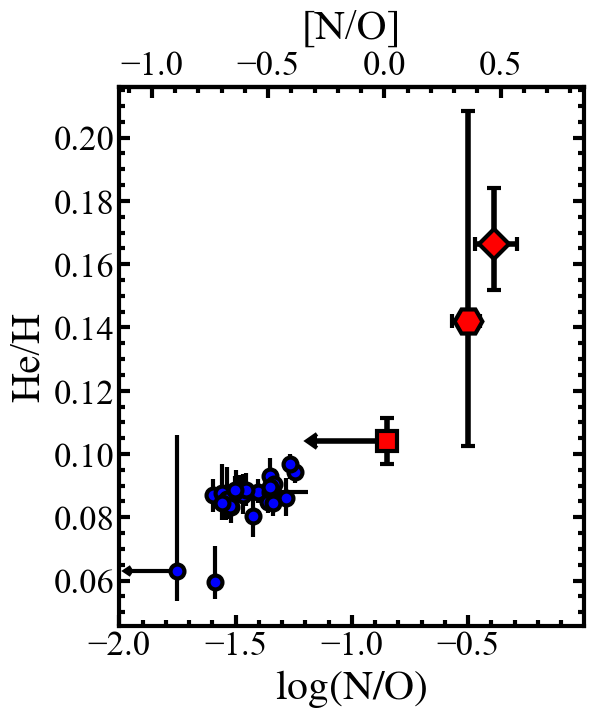}
    \caption{He/H as a function of N/O. The symbols are the same as Figure \ref{fig:HeH-OH}. To avoid overlapping the data points, the point for GLASS150008 is shifted to the left by 0.1 dex.}
    \label{fig:HeH-NO_result}
\end{figure}

\begin{figure*}
    \centering
    \includegraphics[width=1\linewidth]{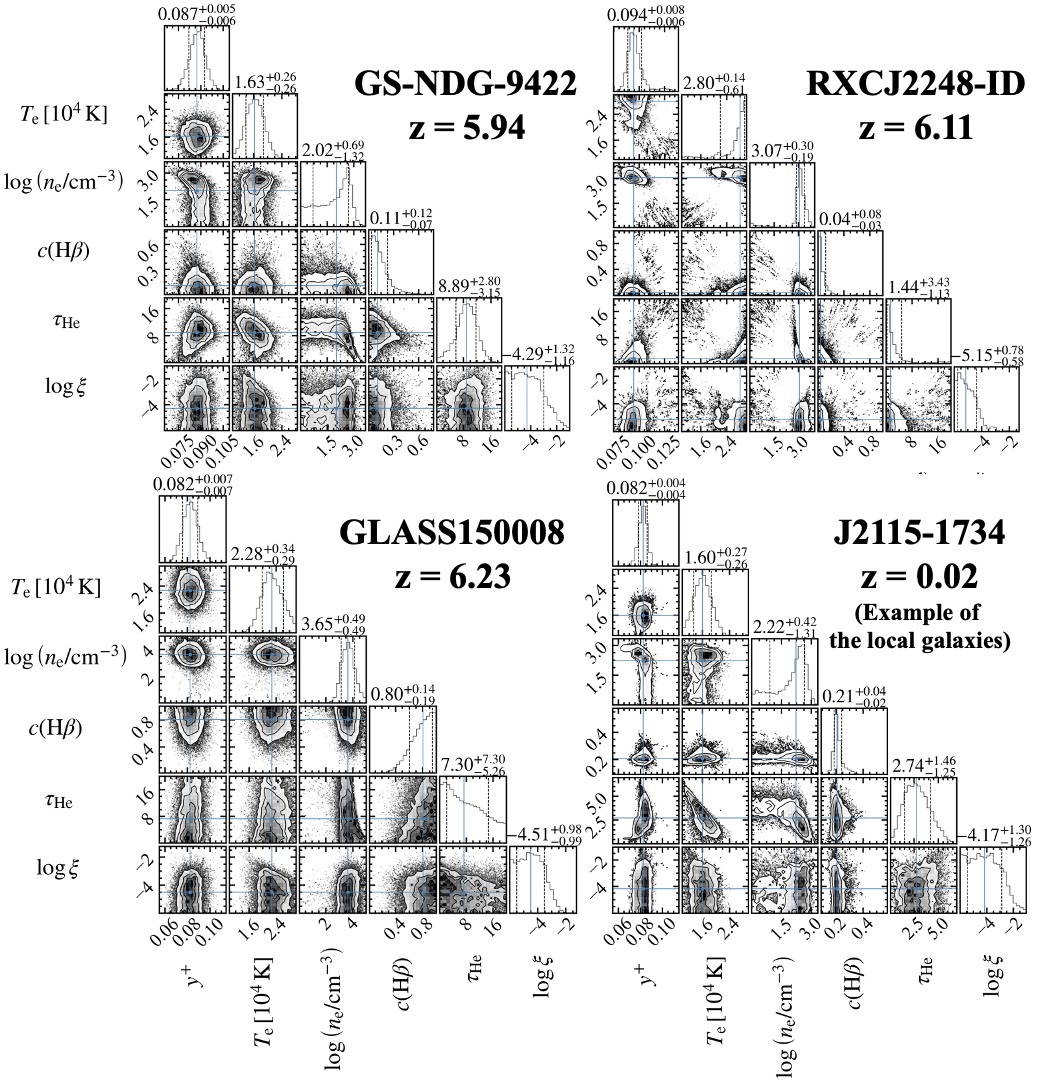}
    \caption{Same as Figure \ref{fig:mcmc}, but with the Gaussian prior on $y^{+}$.}
    \label{fig:mcmc_2}
\end{figure*}

\section{Result} \label{sec:result}
In Figure \ref{fig:HeH-OH}, we show the helium and oxygen abundances of our sample. RXCJ2248-ID and GLASS150008 show the helium abundances that are significantly higher than those of the local galaxies. These galaxies show the large He \textsc{i}/H$\beta$ flux ratios as presented in Figure \ref{fig:fluxratio}. 
In Figure \ref{fig:HeH-OH}, He/H of GS-NDG-9422 derived in this work and by \cite{Cameron+2023} are presented as the red and open squares, respectively.   
The result of \cite{Cameron+2023} are consistent with our result within the 1$\sigma$ level. Note that some local galaxies show He/H lower than the primordial value, although these values are consistent with those derived by \cite{Matsumoto+2022}, who conducted YMCMC analyses with smaller systematics using $\sim$12 emission lines. In Figure \ref{fig:HeH-NO_result}, the He/H and N/O values of our sample galaxies are presented. GLASS150008 and RXCJ2248-ID, which show the helium overabundance in Figure \ref{fig:HeH-OH}, also have large N/O values. We can see a positive correlation between He/H and N/O. These He-rich galaxies introduce significant uncertainty on the primordial helium abundance, which is determined by estimating an intercept of linear fitting in He/H-O/H plane. We suggest that galaxies with large N/O ratios should not be used for the determination of the primordial helium abundance because the N-rich galaxies are likely to have the high He/H ratios (see also \citealt{Izotov+2007}, who suggest the positive correlation between He/H and N/H). 

There may exist the neutral helium in the outer layer of the ionized hydrogen region, which we do not consider at present. If the neutral helium exists, the He/H ratio still increases, which strengthens the result of helium overabundance.

The electron densities of GS-NDG-9422 and GLASS150008 derived from YMCMC are consistent with the upper limits of $\log (n_\mathrm{e}/\mathrm{cm^{-3})}<3$ and $<3.9$ estimated by \cite{Cameron+2023} and \cite{Jones+2023} with the nebular continuum and the C \textsc{iii}]$\lambda\lambda$1907,1909 doublet ratio, respectively. However, the electron density of RXCJ2248-ID derived from YMCMC is lower than $\log (n_\mathrm{e}/\mathrm{cm^{-3})}=4-5$, which are derived with the line ratios of doubly and triply ionized metal lines including C  \textsc{iii}]$\lambda\lambda$1907,1909 \citep{Topping+2024}. 
There is a possibility that the strong He \textsc{i} emission is produced by collisional excitation due to the high electron density.

\begin{deluxetable*}{cccccccc}
    \tablewidth{12pt}
    \tablecaption{Best recovered parameters from MCMC analysis with the Gaussian priors on $y^{+}$} 
    \tablehead{ 
    \colhead{Name} & \colhead{$y^{+}$} & \colhead{$T_\mathrm{e}$} & \colhead{$\log (n_\mathrm{e}/\mathrm{cm^{-3}})$} & \colhead{$c$(H$\beta$)} & \colhead{$\tau_{\rm He}$} & \colhead{$\log\xi$} & \colhead{$\log N_2$} \\
    &&[K]&&&&& \\
    (1)&(2)&(3)&(4)&(5)&(6)&(7)&(8)
    }
    \startdata
    GS-NDG-9422 & $0.087^{+0.005}_{-0.006}$ & $16339^{+2568}_{-2559}$ & $2.02^{+0.69}_{-1.32}$ & $0.11^{+0.12}_{-0.07}$ & $8.89^{+2.80}_{-3.15}$ & $-4.29^{+1.32}_{-1.16}$ & $12.1^{+1.3}_{-2.0}$ \\
    RXCJ2248-ID & $0.094^{+0.008}_{-0.006}$ & $28012^{+1428}_{-6060}$ & $3.06^{+0.30}_{-0.19}$ & $0.04^{+0.08}_{-0.03}$ & $1.44^{+3.43}_{-1.13}$ & $-5.1^{+0.8}_{-0.6}$ & $13.4^{+0.1}_{-0.1}$ \\
    GLASS150008 & $0.082^{+0.007}_{-0.007}$ & $22757^{+3408}_{-2931}$ & $3.65^{+0.49}_{-0.49}$ & $0.80^{+0.14}_{-0.19}$ & $7.30^{+7.30}_{-5.26}$ & $-4.5^{+1.0}_{-1.0}$ & $11.3^{+1.6}_{-1.6}$ \\
    \tableline
    J2115-1734$^\dagger$ & $0.082^{+0.003}_{-0.004}$ & $15992^{+2655}_{-2625}$ & $2.22^{+0.42}_{-1.30}$ & $0.21^{+0.04}_{-0.02}$ & $2.74^{+1.46}_{-1.25}$ & $-4.2^{+1.3}_{-1.26}$ & $11.3^{+1.4}_{-1.6}$ 
    \enddata
    \tablecomments{Same as Table \ref{table:mcmc_recovered_params}, but with the Gaussian priors on $y^{+}$ (see Section \ref{sec:result}). ($\dagger$) One out of the 68 local galaxies is presented as an example. } 
    \label{table:mcmc_recovered_params_2}
\end{deluxetable*}

Including the high electron density scenario, we adopt a Gaussian prior on $y^{+}$ with a central value of 0.08 and standard deviation of 0.01 in the modified YMCMC analyses to examine whether the strong He \textsc{i} emission can be reproduced with the low $y^{+}$. The probability distributions and best-fit parameters are shown in Figure \ref{fig:mcmc_2} and Table \ref{table:mcmc_recovered_params_2}, respectively. For GS-NDG-9422, the best-estimate electron density is $\sim 100\,\mathrm{cm^{-3}}$, while those of RXCJ2248-ID and GLASS150008 are as high as $1,000\,\mathrm{cm^{-3}}$. For GS-NDG-9422 and GLASS150008, these electron densities are consistent with the upper limits of those derived by \cite{Cameron+2023} and \cite{Jones+2023}, respectively. For RXCJ2248-ID, $\log (n_\mathrm{e}/\mathrm{cm^{-3}}) \sim 3$ is still lower than that derived from C \textsc{iii}] doublet by \cite{Topping+2024} by $\sim$2 dex. This may be explained as the distribution of the metal is different from that of helium. If the metal enrichment occurs only within the central region, the metal emission lines suggest the very dense condition. However, in this scenario, helium may exist not only in the central metal-enriched region, but also in the outer pristine gas, which leads to the lower electron density than those of metals.

\begin{figure}
    \centering
    \includegraphics[width=1\linewidth]{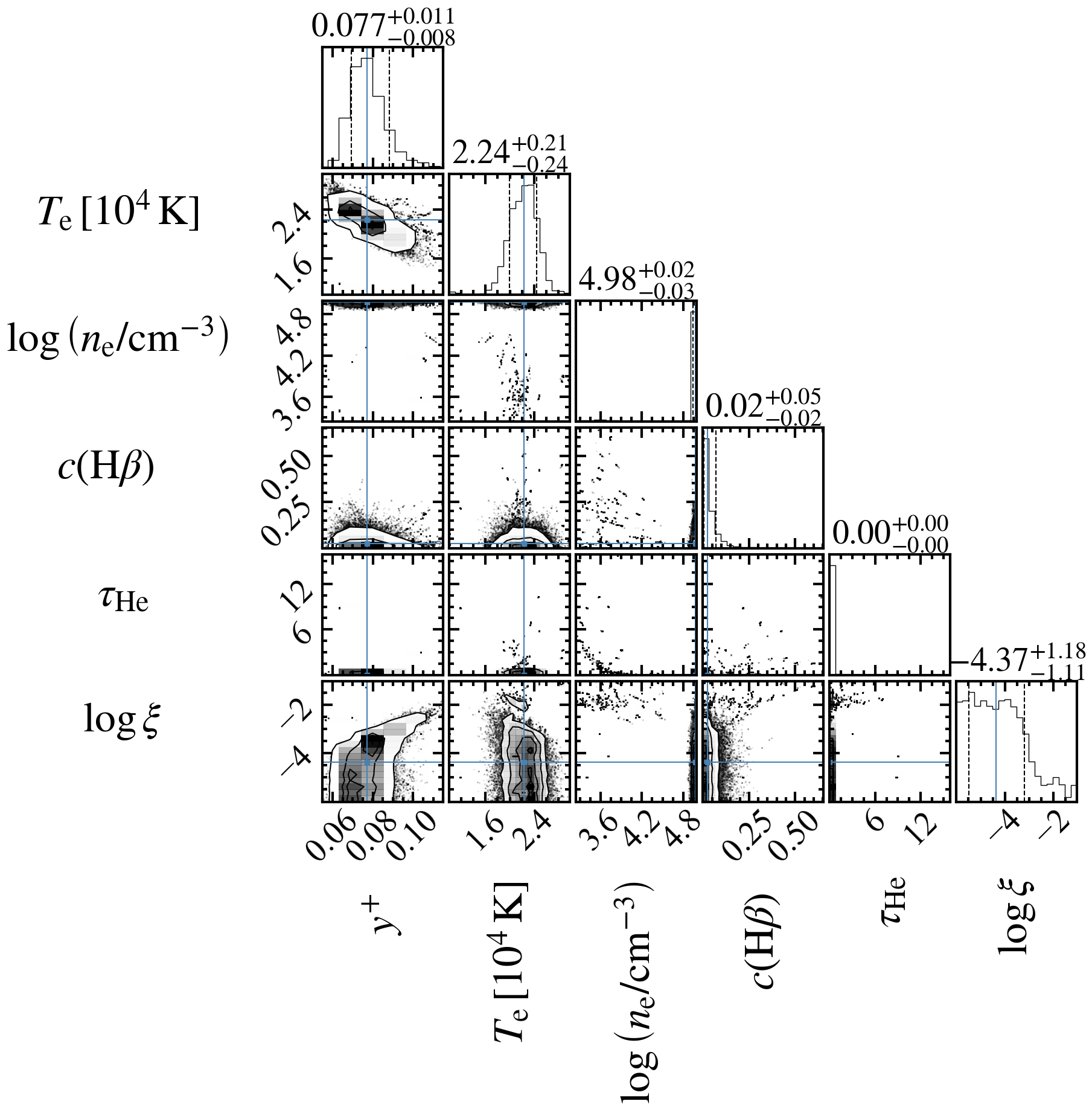}
    \caption{Same as Figure \ref{fig:mcmc}, but for RXCJ2248-ID and with the Gaussian prior on $n_\mathrm{e}.$}
    \label{fig:mcmc_dens_prior}
\end{figure}

\begin{figure*}
    \centering
    \includegraphics[width=1\linewidth]{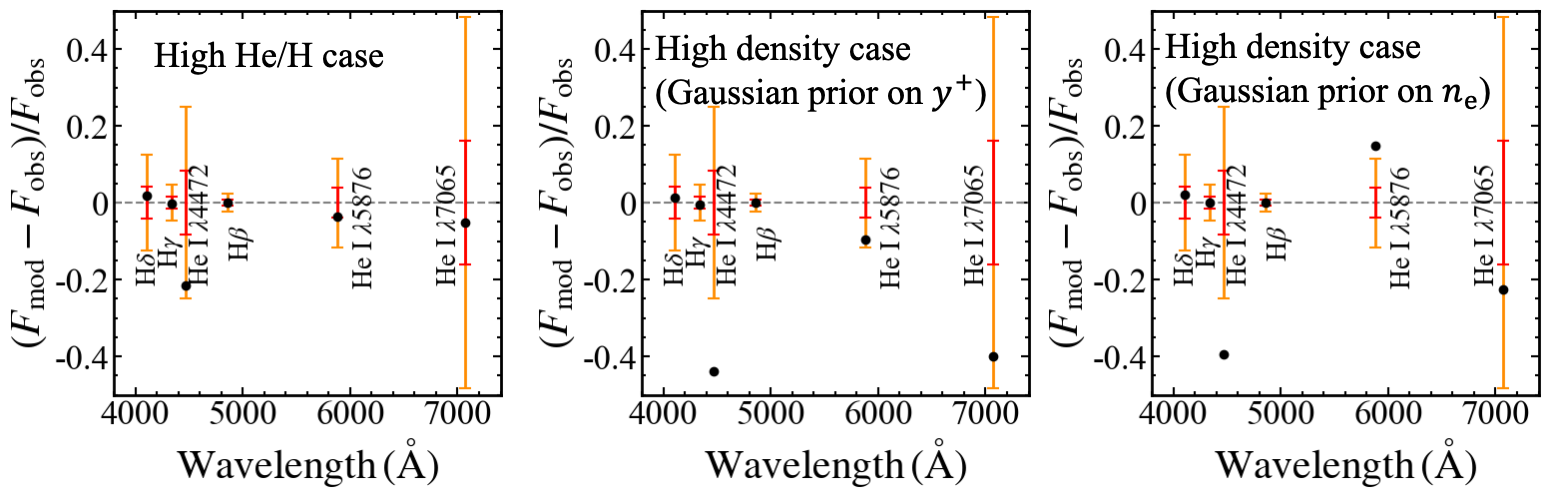}
    \caption{Comparison of the observed flux ratios and those reproduced with modified YMCMC for RXCJ2248-ID. The vertical axes represent the relative error, $(F_\mathrm{mod} - F_\mathrm{obs})/F_\mathrm{obs}$, where $F_\mathrm{mod}$ and $F_\mathrm{obs}$ are the model and observed flux ratios relative to H$\beta$, respectively. The red and orange errorbars denote the 1$\sigma$ and 3$\sigma$ relative errors, respectively. The black dots denote the model flux ratios. From left to right, we present the high He/H case, high density case with the Gaussian prior on $y^+$, and high density case with the Gaussian prior on $n_\mathrm{e}$.}
    \label{fig:match}
\end{figure*}

We also run modified YMCMC for RXCJ2248-ID putting the Gaussian prior on $\log{n_\mathrm{e}}$. We use the Gaussian prior having the central value of 5.04 and standard deviation of 0.06, which is based on the electron density derived from C \textsc{iii}] $\lambda$1907, 1909 doublet \citep{Topping+2024}. The result is shown in Figure \ref{fig:mcmc_dens_prior}. When we constrain $\log (n_\mathrm{e}/\mathrm{cm^{-3})}$ to be $\sim5$, we obtain $y^{+}=0.077$, which is consistent with the typical value of local galaxies. However, the recovery of the flux ratios are even worse (the right panel of Figure \ref{fig:match}).

We also conduct the modified MCMC analysis putting the Gaussian prior on $y^+$ for one of the local dwarf galaxies, J2115-1734 (Figure \ref{fig:mcmc_2}). For the local galaxy, the result is unchanged from the one without the Gaussian prior on $y^{+}$ because the $y^{+}$ value is $\sim 0.08$ even without the Gaussian prior. 

\begin{figure}
    \centering
    \includegraphics[width=1\linewidth]{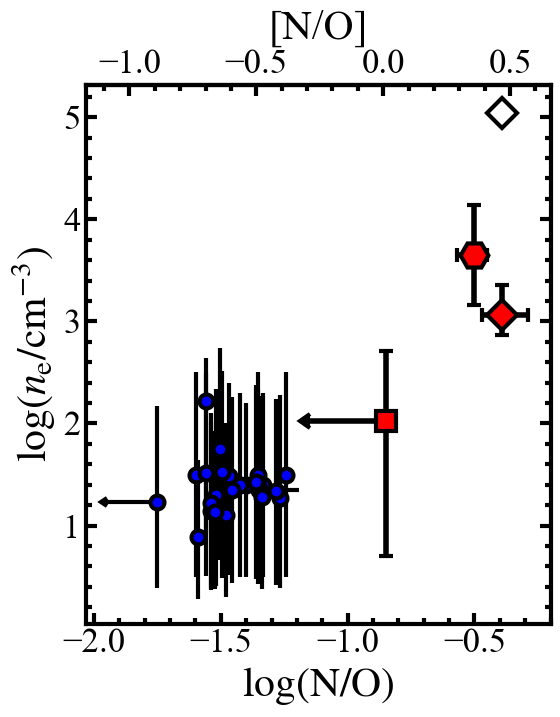}
    \caption{Electron density as a function of N/O. Here $y^{+}$ values are assumed to be consistent with the local values of $y^{+}\sim 0.08$ by putting the Gaussian prior on $y^{+}$. The symbols are the same as Figure \ref{fig:HeH-OH}. The white diamond denotes the electron density for RXCJ2248-ID derived from the C \textsc{iii}] $\lambda$1907,1909 doublet \citep{Topping+2024}. To avoid overlapping the data points, the red hexagon is shifted to the left by 0.1 dex.}
    \label{fig:NO-ne}
\end{figure}

In Figure \ref{fig:NO-ne}, the relation between N/O and $n_e$ is presented. Under the assumption of the low $y^{+}$, we can see the positive correlation between N/O and $n_e$, which may suggest that the dense situation leads to the excessive enhancement of nitrogen. The difference in $n_\mathrm{e}$ between the high-$z$ and local galaxies may be originated from the compact morphology in high-$z$ galaxies, as suggested by \cite{Isobe+2023_ne}.

Hereafter the case where the density is low and He/H is high is referred to as high He/H case, and the case where the density is high and He/H is low is referred to as high density case. In Figure \ref{fig:match}, we show the comparison of the observed and recovered flux ratios for RXCJ2248-ID. In the high He/H case, the observed flux ratios are reproduced better than those in the high density case. However, in the high He/H case, it is difficult to explain the discrepancy in the electron densities derived with modified YMCMC and C \textsc{iii}] doublet. It appears that both of the cases do not fully explain the observed quantities for RXCJ2248-ID. In either cases, however, the high-$z$ galaxies in our sample are quite different from the local galaxies, whose He/H and $n_\mathrm{e}$ values derived in the same manner show significantly lower than those of the high-$z$ galaxies.

\section{Discussion} \label{sec:discussion}
In the previous section, we find that the strong He \textsc{i} emission is achieved by either the high He/H ratios or the high electron densities. In the following sections, assuming the high He/H case, we compare the observed chemical compositions of He/H, O/H, and N/O with the chemical evolution models.

\subsection{He/H and O/H} \label{sec:discussion_heh}
\subsubsection{Comparison with the CCSNe Model} \label{sec:CCSN}
We first compare the observed He/H and O/H ratios with a standard core-collapse supernovae (CCSNe) model. We use the CCSNe model developed by \cite{Watanabe+2023}, who calculate chemical evolution of galaxies by utilizing the explosive nucleosynthesis code developed by \cite{Tominaga+2007}. In this model, stars in a mass range of $9\,M_\odot \leq M \leq 40\,M_\odot$ are instantaneously created according to Kroupa IMF \citep{Kroupa_2001}. At the end of the lifetime, these stars cause CCSNe and eject gas. The yield of CCSNe is taken from \cite{Nomoto+2013}. For the details of the model, see \cite{Watanabe+2023}. In this model, the ejecta have abundances of $\mathrm{He/H} \sim 0.20$ and $12 + \log \mathrm{(O/H)} \sim 10$. Because the high-$z$ galaxies in our sample have the oxygen abundances of $12 + \log \mathrm{(O/H)} \sim 7.4-7.7$, the ejecta need to be diluted by primordial gas in the interstellar medium (ISM). The observed emission lines come from this mixture of ejecta and primordial gas. We thus calculate the He/H and O/H abundance ratios in the mixed gas that are given by

\begin{equation}\label{eq:mixing}
\begin{split}
    & \textrm{He/H} = r_\mathrm{mix}(\textrm{He/H})_\textrm{e} + (1 - r_\mathrm{mix}) (\textrm{He/H})_\textrm{p} \\
    & \textrm{O/H} = r_\mathrm{mix}(\textrm{O/H})_\textrm{e} + (1 - r_\mathrm{mix}) (\textrm{O/H})_\textrm{p}, \\
\end{split}
\end{equation}
where the subscripts e and p stand for `ejecta' and `primordial gas', respectively, and $r_\mathrm{mix}$ is a mixing parameter which satisfies $0 < r_\mathrm{mix} \leq 1$. $r_\mathrm{mix}$ is defined so that $r_\mathrm{mix}=0$ corresponds to the primordial gas and $r_\mathrm{mix}=1$ corresponds to the pure ejecta. 

In Figure \ref{fig:HeH-OH_model}(a), we compare our observational results with the CCSNe model. Because the oxygen abundance is rapidly enhanced by the CCSNe ejecta, $r_\mathrm{mix}$ needs to be smaller than 0.01 to fall within the observed O/H range. However, the small $r_\mathrm{mix}$ reduces the He/H to almost the primordial value of 0.08, which indicates that the CCSN model is unlikely to reproduce the observed high He/H and low O/H. The same conclusion was also reached by \cite{Fukushima+2024} using a different one-zone chemical evolution model.

\subsubsection{Comparison with the Other Models} \label{sec:other_models}
\begin{figure*}
    \centering
    \includegraphics[width=1\linewidth]{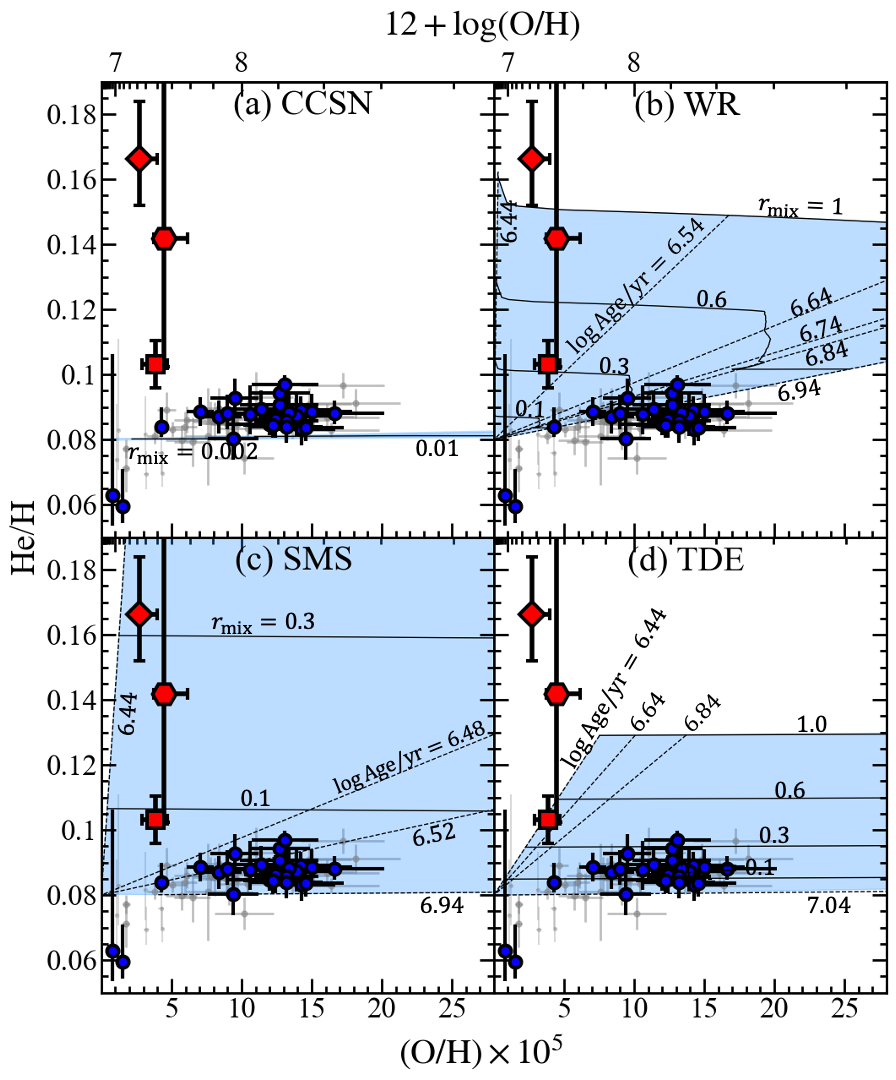}
    \caption{Comparison with chemical evolution models calculated by \cite{Watanabe+2023} and Watanabe et al. (in prep.). The symbols are the same as Figure \ref{fig:HeH-OH}. The blue shaded regions show the regions that are reproduced by the models with $0\leq r_\mathrm{mix}\leq1$. (a): CCSN model. The black lines indicate the models with $r_\mathrm{mix}=0.002$ and $0.1$. Models with $r_\mathrm{mix}\gtrsim 0.1$ extend beyond the range of the figure due to the rapid enhancement of oxygen. (b)-(d): WR star, SMS, and TDE models. The black solid lines indicate the model tracks with fixed $r_\mathrm{mix}$ values. The black dashed lines present the models with the fixed values of the log of the age (yr). Here, $f_\mathrm{SN} = 0.005$ is assumed. 
    Note that $r_\mathrm{mix}=0$ corresponds to the point $(\mathrm{O/H})\times 10^5 = 0$, $\mathrm{He/H}=0.08$.}
    \label{fig:HeH-OH_model}
\end{figure*}

We then explore another physical origin of the helium  overabundances. The high He/H may be explained by gas produced in the CNO cycle, which occurs in an outer layer of massive stars and produce a large amount of helium from hydrogen. If the hydrogen burning layer is ejected into the ISM, the He/H value may become high. In CNO cycle equilibrium, a large amount of nitrogen is accumulated compared to oxygen, which also enhances N/O \citep{Isobe+2023c}. In previous studies, models including Wolf-Rayet (WR) stars, supermassive stars (SMSs), and tidal disruption events (TDEs) are presented to explain the anomalous chemical abundances of high-$z$ galaxies such as GN-z11 \citep{Isobe+2023c, Charbonnel+2023, Nagele_Umeda_2023, Watanabe+2023, Senchyna+2023, Cameron+2023a}. We thus incorporate the yields of the WR star, SMS, and TDE into the model. We use chemical evolution models of WR stars developed by \cite{Watanabe+2023} and of SMSs and TDEs developed by Watanabe et al. in (prep.). In these models, stars in a mass range of $9\,M_\odot \leq M \leq 25\,M_\odot$ cause CCSNe at the end of the lifetimes. On the other hand, stars in a mass range of $M > 25\,M_\odot$ end up in CCSNe at a probability $f_\mathrm{SN}$ and direct collapses (DCs) at a probability $1-f_\mathrm{SN}$. The yields of each model are shown in Table \ref{table:yield}.
We allow the mixing that is described as $r_\mathrm{mix}$ in Equation \eqref{eq:mixing}. In Figure \ref{fig:HeH-OH_model}(b), (c), and (d), we compare the observational results with the WR star, SMS, and TDE models, respectively. The WR star model explains the high He/H of the ISM in the three high-$z$ galaxies within the $\sim 1\sigma$ levels. The SMS model also reproduces the He/H and O/H of the ISM in all of the three high-$z$ galaxies. The TDE model explains the chemical compositions of GS-NDG-9422 and GLASS150008 within the 1$\sigma$ level, while that of RXCJ2248-ID is not reproduced by the TDE model. 

\begin{deluxetable*}{ccccccc}
    \tablewidth{12pt}
    \tablecaption{Yields of the models}
    \tablehead{ 
    \colhead{Model} & \colhead{Mass [$M_\odot$]} & \colhead{$^1$H [$M_\odot$]} & \colhead{$^4$He [$M_\odot$]} & \colhead{$^{14}$N [$M_\odot$]} & \colhead{$^{16}$O [$M_\odot$]} & Reference
    }
        \startdata
        WR$^\dagger$ & 13 & 4.52 & 3.89 & 0.13 & 1.51 & \\
           & 15 & 5.02 & 4.51 & 0.15 & 1.82 &  \\
           & 20 & 6.73 & 6.20 & $1.18\times10^{-2}$ & 2.94 &  \\
           & 25 & 7.81 & 5.29 & $1.66\times10^{-2}$ & 5.93 &  \\
           & 30 & 8.94 & 4.00 & $1.69\times10^{-4}$ & $1.55\times10^{-4}$ & \cite{Limongi+2018} \\
           & 40 & 11.1 & 4.38 & $1.90\times10^{-4}$ & $1.99\times10^{-4}$ &  \\
           & 60 & 15.5 & 6.37 & $2.91\times10^{-4}$ & $2.53\times10^{4}$ &  \\
           & 80 & 17.5 & 9.71 & $2.14\times10^{-2}$ & $5.80\times10^{-3}$ &  \\
           & 120& 22.6 & 14.7 & $1.17\times10^{-3}$ & $2.66\times10^{-4}$ &  \\
        \tableline
        SMS & $10^5$ & $4.19\times10^4$ & $5.8\times10^4$ & 81.4 & 0.884 &  \cite{Nagele_Umeda_2023} \\
        \tableline
        TDE & 13 & 4.61 & 1.79 & $5.33\times10^{-3}$ & $1.12\times10^{-2}$ &  \\
            & 15 & 5.21 & 2.20 & $1.63\times10^{-3}$ & $3.13\times10^{-3}$ &  \\
            & 18 & 5.66 & 3.15 & $2.53\times10^{-3}$ & $3.17\times10^{-3}$ &  \\
            & 20 & 6.41 & 3.28 & $2.66\times10^{-3}$ & $3.50\times10^{-3}$ & Watanabe et al. (in prep.) \\
            & 25 & 7.61 & 4.23 & $3.68\times10^{-3}$ & $4.02\times10^{-3}$ &  \\
            & 30 & 8.75 & 5.27 & $4.71\times10^{-3}$ & $4.35\times10^{-3}$ &  \\
            & 40 & 10.5 & 7.30 & $6.60\times10^{-3}$ & $4.88\times10^{-3}$ &  
        \enddata
    \tablecomments{($\dagger$) Metallicity of $\mathrm{[Fe/H]}=-3$ and an initial rotation velocity of $v=300 \, \mathrm{km \, s^{-1}}$ is assumed.} 
    \label{table:yield}
\end{deluxetable*}

\begin{figure*}
    \centering
    \includegraphics[width=1\linewidth]{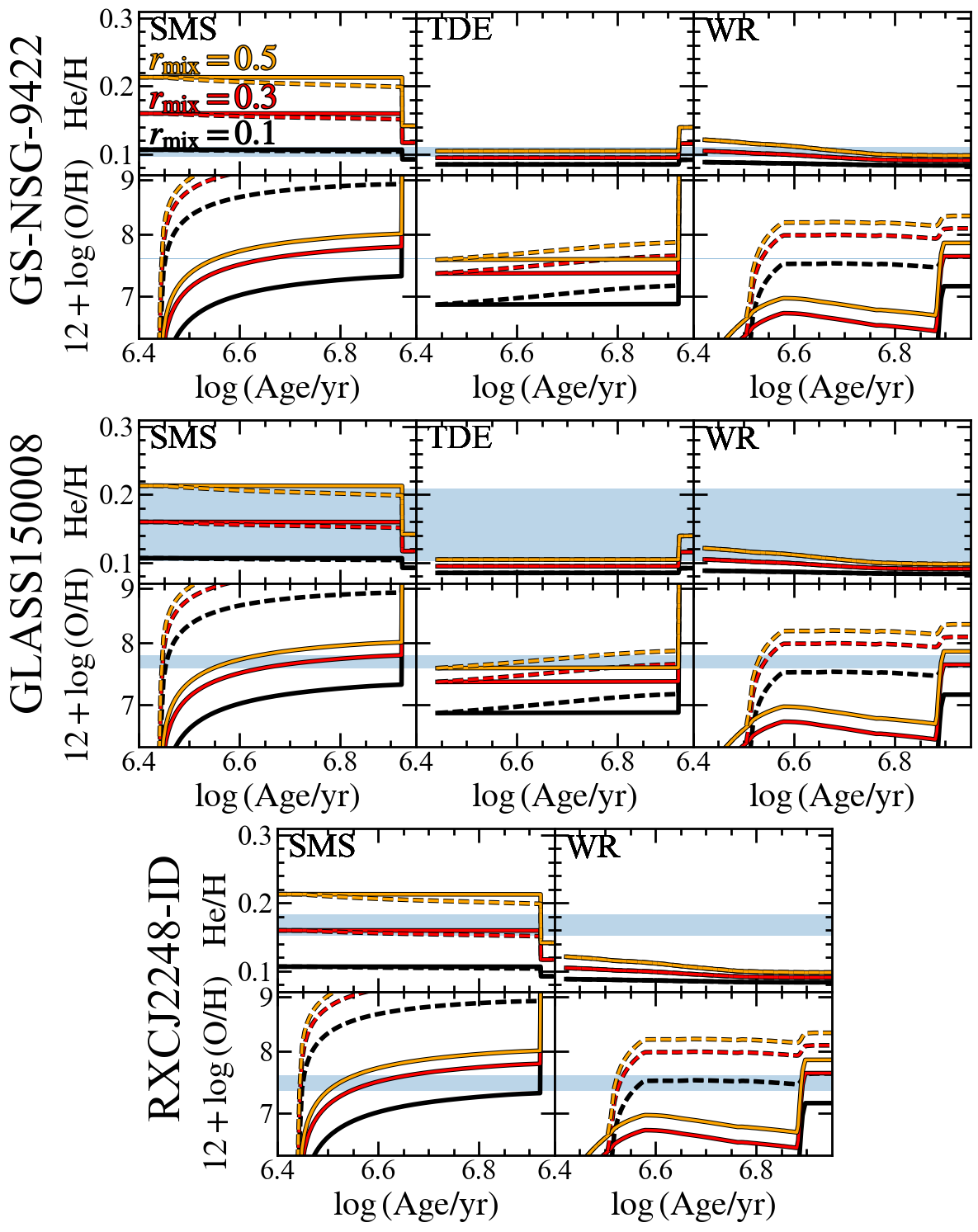}
    \caption{He/H and O/H as a function of age. The black, red and orange lines represent the models with mixing parameters of $r_\mathrm{mix}=0.1$, $0.3$, and $0.5$, respectively. The solid and dashed lines denote the models with $f_\mathrm{SN}$ values of $0.0001$ and $0.005$, respectively. The blue shaded regions in the top, middle, and bottom panels indicate the 1$\sigma$ areas of observed abundance ratios of GS-NDG-9422, RXCJ2248-ID, and GLASS150008, respectively. 
    }
    \label{fig:time_evolution}
\end{figure*}

\begin{figure*}
    \centering
    \includegraphics[width=1\linewidth]{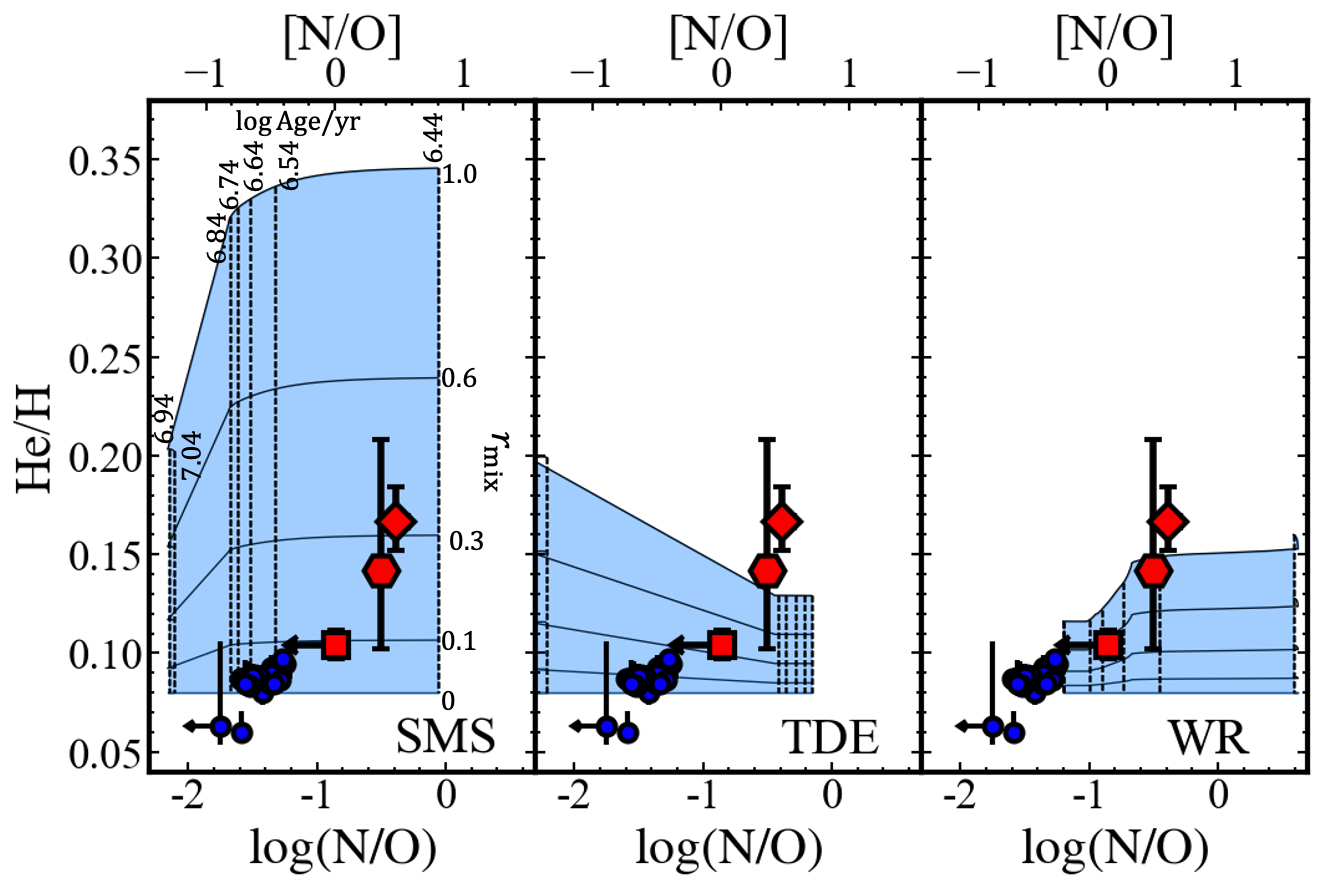}
    \caption{Same as Figure \ref{fig:HeH-OH_model}, but for He/H as a function of N/O. From left to right, SMS, TDE, and WR star models are shown. The symbols are the same as Figure \ref{fig:HeH-OH}. The blue shaded regions represent the regions that are reproduced by the models. The black solid lines indicate the trajectories of each model with fixed mixing parameters shown beside the lines. The black dashed lines represent the model tracks with the fixed values of the log of the age. Here, $f_\mathrm{SN} = 0.005$ is assumed. Since the grids in the three panels are drawn at the same mixing parameters and stellar ages, the values are only shown in the left panel.}
    \label{fig:HeH-NO_model}
\end{figure*}
In Figure \ref{fig:time_evolution}, we show the He/H and O/H of the models as a function of age. For RXCJ2248-ID, only the SMS and WR star model is presented since the TDE model does not explain the He/H and O/H values of RXCJ2248-ID. We calculate the chemical evolution at $r_\mathrm{mix}=0.1-0.5$ and $f_\mathrm{SN}=0.0001-0.005$. At $\log \mathrm{(Age/yr)} \sim 6.9$, O/H rapidly increases because the stars with $M < 25 M_\odot$ start to cause CCSNe.
For GS-NDG-9422, the SMS model reproduces the He/H and O/H ratios within the 1$\sigma$ levels for $r_\mathrm{mix}=0.1$ and $f_\mathrm{SN}=0.005$, while the O/H value is reproduced within the very short timescale because of the rapid oxygen enhancement by CCSNe. On the other hand, the TDE model with $r_\mathrm{mix}=0.5$ and $f_\mathrm{SN}=0.0001$ reproduces He/H and O/H within the longer timescale ($\log (\mathrm{Age/yr})=6.45-6.9$) due to the moderate oxygen enrichment by the TDE. The WR model with $r_\mathrm{mix}=0.3$ and $f_\mathrm{SN}$ value between 0.0001 and 0.005 is likely to reproduce the He/H and O/H of GS-NDG-9422. For GLASS150008, the SMS model reproduces the He/H and O/H values with all of the parameter sets shown in Figure \ref{fig:time_evolution} except for $r_\mathrm{mix}=0.1$ and $f_\mathrm{SN}=0.0001$, 
while the TDE models with $r_\mathrm{mix}\sim0.5$ explain the He/H and O/H within the $1\sigma$ level. The WR models with $r_\mathrm{mix}=0.3-0.5$ and $f_\mathrm{SN}\sim0.005$ are likely to reproduce the He/H and O/H on GLASS150008. For RXCJ2248-ID, the SMS model reproduces the observed He/H and O/H within the $\sim1\sigma$ levels for both of $f_\mathrm{SN}=0.0001$ and 0.005 if $r_\mathrm{mix}=0.3$ is assumed, while the WR model requires larger $r_\mathrm{mix}$ values of $\sim1$.

In all cases explained above, the large fraction of massive stars need to cause DCs ($f_\mathrm{SN}\lesssim0.01$) to explain the observed O/H within the sufficiently long timescale. This is consistent with the result of \cite{Watanabe+2023}, who suggest that most of WR stars need to cause DCs to explain the observed large N/O in GN-z11. However, this result of low $f_\mathrm{SN}$ depends on the assumed IMF and star-formation timescale, which need to be further investigated.

Another possible source of the anomalous chemical abundances is very massive stars (VMSs; $100-1000\,M_\odot$), whose stellar winds may be responsible for N/O, He/H, and He \textsc{ii} emission \citep{Vink_2023}. Although the VMSs are not incorporated in this work, these stars with intermediate masses between WR stars and SMSs could be an alternative origin of the observed anomalous chemical abundances.

\subsection{He/H and N/O} \label{sec:discussion_Nitrogen}
In Figure \ref{fig:HeH-NO_model}, we compare the N/O ratios with the same models as presented in Section \ref{sec:other_models}. In the SMS model, $r_\mathrm{mix} \sim 0.3$ (0.1) reproduces the chemical abundances of RXCJ2248-ID and GLASS150008 (GS-NDG-9422), which are consistent with the result presented in Figure \ref{fig:HeH-OH_model}. In the TDE and WR star models, chemical abundances of GS-NDG-9422 and GLASS150008 are reproduced at various values of $r_\mathrm{mix}$, while those of RXCJ2248-ID are not explained by any $r_\mathrm{mix}$ within the 1$\sigma$ level. These results suggest that the high He/H and N/O is achieved by the gas produced by the CNO-cycle, which supplies a large amount of both He and N. 

These He- and N-rich situations are not found in the local dwarf galaxies. This may arise from two reasons. First, the CNO-cycled gas is sufficiently diluted by the primordial gas and ejecta of the CCSNe in the local dwarf galaxies, which leads to the low He/H and N/O ratios. Second reason may be massive star-formation in the high-$z$ galaxies, which produces more massive stars such as WR stars, VMSs, and SMSs ejecting CNO-cycled gas. Given that the He- and N-rich galaxies may exist in the high-$z$ universe but not in the local universe, it is implied that the massive star-formation occur in these galaxies with anomalous chemical abundances.

\section{Summary} \label{sec:summary}
In this paper, using the MCMC algorithm, we derive the He/H abundance ratios of 3 high-$z$ galaxies and compare them with the chemical evolution models. Our main results are summarized below.
\begin{itemize}
    \item Among the 5 galaxies at $z\gtrsim6$ that have the N/O constraints, we select 3 galaxies whose optical helium and hydrogen lines necessary for the He/H estimates are covered by JWST NIRSpec. We also select the local dwarf galaxies as the control sample. We find the flux ratios of He \textsc{i} to H$\beta$ in the 3 high-$z$ galaxies are comparable to or significantly larger than those in the local galaxies (Figure \ref{fig:fluxratio}).
    
    \item We obtain the He/H with modified YMCMC code, which is the modified version of the YMCMC code \citep{Hsyu+2020} developed for the primordial helium abundance determination. The He/H abundance ratio of the high-$z$ galaxies are higher than those of the local galaxies (Figure \ref{fig:HeH-OH}). We also derive He/H for the local galaxies as a control sample using the same helium and hydrogen lines. These local galaxies show the low helium abundances, meaning that the limited number of input emission lines is not the cause of the helium overabundances. We also compare our sample galaxies in the He/H-N/O plane (Figure \ref{fig:HeH-NO_result}). We find that He/H has the positive correlation with N/O. This may be caused by the gas produced by the CNO-cycle, which enhances both He and N.

    \item There is the possibility that the strong He \textsc{i} emission is achieved by the excessive collisional excitation caused by the high electron density, not by the high He/H ratio. We rerun modified YMCMC with the Gaussian prior on $y^{+}$ with the central value of 0.08, which is nearly the primordial helium abundance. We find that if the high-$z$ galaxies in our sample have the He/H ratios comparable to the local galaxies, these high-$z$ galaxies have the electron densities significantly higher than those of the local galaxies (Figure \ref{fig:NO-ne}). In this case, we also find the positive correlation between N/O and $n_\mathrm{e}$, implying that the nitrogen enrichment may be connected with the dense situation.

    \item We also conduct analyses with the modified YMCMC for the control sample of 68 local dwarf galaxies in the same manner. These local galaxies show the low He/H and $n_e$ values, which indicate that in either of the high He/H and high density case, the high-$z$ galaxies in our sample are different from the local galaxies and these differences are not caused by systematics.
    
    \item Assuming the high-$z$ galaxies in our sample have the high helium abundance, we compare the He/H and O/H abundance ratios with the standard chemical enrichment by the CCSNe (Figure \ref{fig:HeH-OH_model} (a)). The helium overabundance in the high-$z$ galaxies are unlikely to be explained by the abundance of the mixed gas of the CCSNe ejecta and primordial gas. We then introduce the WR star, TDE, and SMS models, which eject He-rich CNO-cycled gas into the ISM, to explain the high He/H values in the high-$z$ galaxies (Figure \ref{fig:HeH-OH_model} (b), (c), and (d), respectively). The WR star and SMS models reproduce the He/H and O/H in the three high-$z$ galaxies. The TDE model explains the He/H and O/H values observed in GS-NDG-9422 and GLASS150008, while those of RXCJ2248-ID are not reproduced in this model. Comparing the time evolution of He/H and O/H of the models with the observational results (Figure \ref{fig:time_evolution}), we find that 
    low $f_\mathrm{SN}$ values ($\lesssim$ 1 \%) are needed to explain the observed abundances in sufficiently long timescales.

    \item We also compare the He/H and N/O values with the models (Figure \ref{fig:HeH-NO_model}). We find that the He/H and N/O ratios of GS-NDG-9422 and GLASS150008 are reproduced by the SMS, TDE, and WR star models, while those of RXCJ2248-ID are reproduced simultaneously only by the SMS model with $r_\mathrm{mix} \sim 0.3$ within the $1\sigma$ level.
\end{itemize}

\section*{Acknowledgement}
We are thankful to Daniel P. Stark, Ricardo Amorin, and Siwei Zou for helpful discussions. This work is based on the observations with the NASA/ESA/CSA James Webb Space Telescope associated with programs of GTO-1210 (JADES) and GO-2478. We are grateful to GTO-1210 and GO-2478 teams led by Nora Luetzgendorf and Daniel P. Stark, respectively. The JADES data presented in this article were obtained from the Mikulski Archive for Space Telescopes (MAST) at the Space Telescope Science Institute. The specific observations analyzed can be accessed via \dataset[doi:10.17909/8tdj-8n28]{https://doi.org/10.17909/8tdj-8n28}. This publication is based upon work supported by the World Premier International Research Center Initiative (WPI Initiative), MEXT, Japan, KAKENHI (20H00180, 21H04467, and 21H04489) through Japan Society for the Promotion of Science, and JST FOREST Program (JP-MJFR202Z). This work was supported by the joint research program of the Institute for Cosmic Ray Research (ICRR), University of Tokyo.

\vspace{5mm}

\software{astropy \citep{astropy_2013, astropy_2018, astropy_2022}, NumPy \citep{Harris+2020_numpy}, matplotlib \citep{Hunter+2007_matplotlib}, SciPy \citep{Virtanen+2020_scipy}, corner \citep{Foreman-Mackey_2016_corner}, emcee \citep{Foreman-Mackey_2013_emcee}, and YMCMC \citep{Hsyu+2020}
          }

\bibliography{reference}{}
\bibliographystyle{aasjournal}

\end{document}